\documentclass[final]{aipproc}

\layoutstyle{8x11single}

\begin{document}

\title{Bottom and charmed hadron spectroscopy from lattice QCD}


\author{Randy Lewis}{address={Department of Physics and Astronomy,
                            York University, Toronto, Ontario, Canada M3J 1P3}}

\begin{abstract}
A survey of recent lattice QCD simulations for the mass spectrum of bottom and
charmed hadrons is presented.
\end{abstract}

\maketitle

%%%%%%%%%%%%%%%%%%%%%%%%%%%%%%%%%%%%%%%%%%%%
%% MAINMATTER
%%%%%%%%%%%%%%%%%%%%%%%%%%%%%%%%%%%%%%%%%%%%

\section{Introduction}

During the three years since the previous meson-nucleon (MENU) conference,
there have
been many lattice QCD computations of bottom and charmed hadron masses.
These studies must handle challenges due to the physical quark masses being
spread out relative to the natural lattice scales, $\lambda_{\rm QCD}$ and
$1/a$ (the ultraviolet cutoff):
\begin{equation}
m_u\approx m_d ~~~<~~~ m_s\sim\Lambda_{\rm QCD} ~~~<~~~ m_c ~~~<~~~ \frac{1}{a}
~~~<~~~ m_b~.
\end{equation}

\section{Bottom baryons}

During the past three years, the CDF and D0 collaborations have published
first observations of the $\Sigma_b$\cite{CDF}, $\Sigma_b^*$\cite{CDF},
$\Xi_b$\cite{D0,CDF2} and $\Omega_b$\cite{OmegabD0,OmegabCDF}
baryons.  The two results for the $\Omega_b$ are not consistent with one
another.

Various lattice QCD groups have computed the mass spectrum using
different b-quark actions ({\em NRQCD}\cite{LW,MDLW},
{\em Fermilab}\cite{NG}, {\em static}\cite{BHLLS,DLW,LCMO}),
different light-quark actions
({\em nonperturbatively-tuned clover}\cite{LW}, {\em improved
staggered}\cite{NG},
{\em chirally-improved}\cite{BHLLS},
{\em domain wall}\cite{MDLW,DLW}, {\em improved-staggered
sea with domain wall valence}\cite{LCMO}) and different gauge actions
({\em Iwasaki}\cite{LW,MDLW,DLW}, {\em one-loop
Symanzik/L\"uscher-Weisz}\cite{NG,BHLLS,LCMO}).
Any action with the correct continuum limit is a valid approach, and these
complementary studies provide valuable information about systematic
uncertainties at present lattice spacings and volumes.
Chiral extrapolations toward the physical u and d quark masses are also a
source of systematic uncertainty, and the recent bottom baryon studies reach
a minimum pion mass near 275 MeV\cite{DLW}, 290 MeV\cite{NG,LCMO},
331 MeV\cite{MDLW}, 461 MeV\cite{BHLLS} and 600 MeV\cite{LW}.

\begin{figure}\label{bottombaryons}
\includegraphics[width=17cm,trim=130 510 20 120,clip=true]{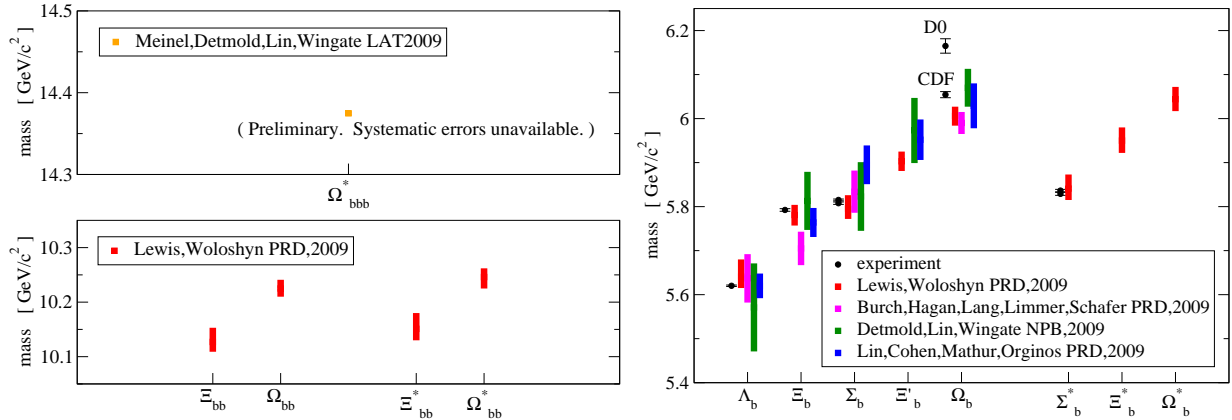}
\caption{Bottom baryon spectrum from lattice QCD from
\cite{LW,MDLW,BHLLS,DLW,LCMO}.  Not shown are
preliminary results from \cite{MDLW} for singly- and doubly-b baryons,
valuable data from \cite{NG} for mass differences among singly- and doubly-b
baryons, and the prediction for $M_{\Omega_{bbb}}$ given in equation
(\ref{eqMeinel}).}
\end{figure}

Figure~\ref{bottombaryons} conveys an impression of some results from these
recent
studies of bottom baryons, but of course the original publications provide
many additional insights.  For example, \cite{NG} provides
a discussion of systematically-large splittings
for $(\Lambda_b,\Xi_b)$ versus $(\Sigma_b,\Xi_b^\prime,\Omega_b)$
that arise with a staggered lattice action due to the
inability to separate $J^p=\frac{1}{2}^+$
and $\frac{3}{2}^+$ states using the spin projection operators.

It is clear from figure \ref{bottombaryons} that lattice simulations are
consistent with the CDF measurement of the $\Omega_b$ mass as opposed to the D0
result.  Lattice findings also agree with experiment for all other available
masses, and lattice predictions are in place for comparison to future
experimental discoveries.

Two related preprints have appeared after the MENU 2010 conference.
Wagner and Wiese\cite{WW} have computed the baryon spectrum in the limit of
a static bottom quark, where the twisted mass action was used for u,d quarks.
Meinel\cite{Meinel} has produced a
preprint that is focused on the mass of the $\Omega_{bbb}$.  Using NRQCD
bottom quarks on dynamical lattice configurations, he arrives at
\begin{equation}\label{eqMeinel}
M_{\Omega_{bbb}} = 14.371\pm0.004_{stat}\pm0.011_{syst}\pm0.001_{exp} {\rm ~GeV}
\end{equation}
where the third uncertainty arises from required experimental input.
This mass with excusively heavy valence quarks can be of significant value for
testing the content of theoretical models.

\section{Charmed baryons}

In contrast to $m_b$, the charmed quark mass is smaller than the
standard lattice cutoff
so neither NRQCD nor static quark actions are generally useful.
Two lattice groups\cite{NG,LLOW} have recently used the Fermilab action for
the charmed
quark and gauge configurations containing staggered u,d,s sea quarks to obtain
charmed baryon masses.  Liu, Lin, Orginos and Walker-Loud\cite{LLOW} used domain
wall valence u,d,s quarks while Na and Gottlieb\cite{NG} used the staggered
action.  The lightest u,d masses correspond to $m_\pi=290$ MeV.

\begin{figure}\label{LLOWfig14}
\includegraphics[width=9.7cm,trim=0 15 0 0,clip=true]{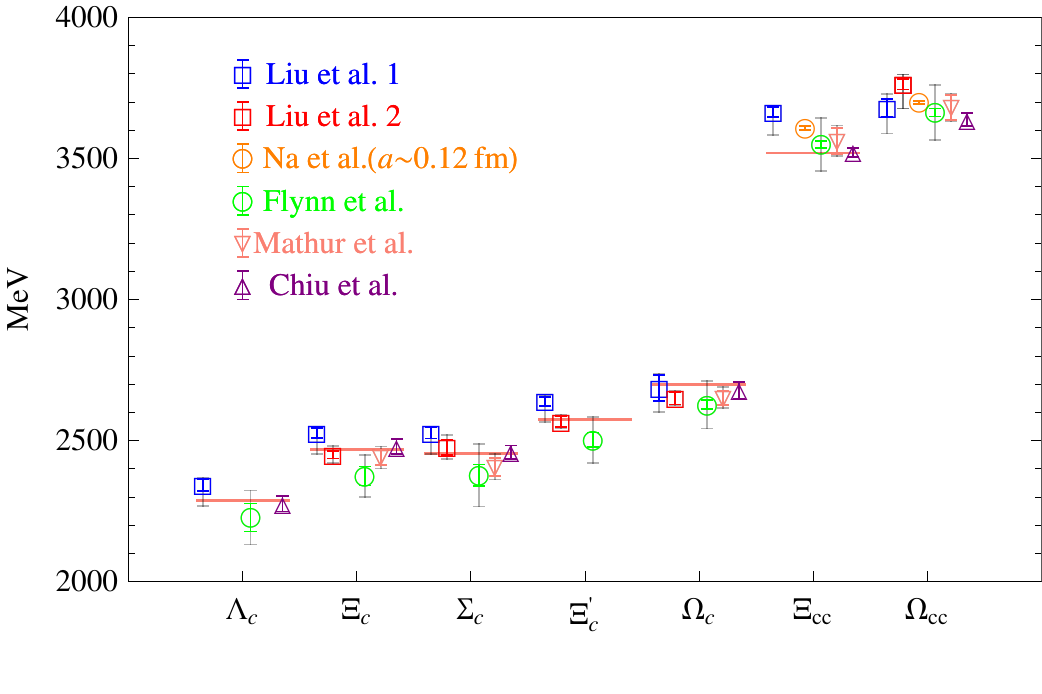}
\caption{Charmed baryon spectrum from lattice QCD\cite{NG,LLOW}
compared to older quenched results from \cite{MLW,FMT,CH}.
Not shown are results in \cite{NG} for mass
differences, including $\Xi_{cc}^*$ and $\Omega_{cc}^*$.  This graph has been
taken directly from \cite{LLOW}.}
\end{figure}

Some results are shown in figure \ref{LLOWfig14}, which was taken directly from
\cite{LLOW}.  It is interesting to see the consistency of the new results with
the older simulations\cite{MLW,FMT,CH} which had relied on the quenched
approximation.

\section{Bottom mesons}

\begin{figure}\label{bottommesons}
\includegraphics[width=11cm,trim=0 50 0 70,clip=true]{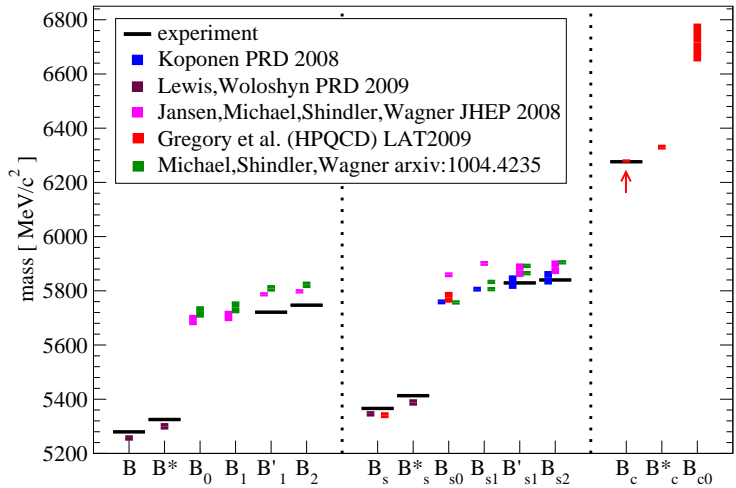}
\caption{Bottom meson spectrum from lattice QCD\cite{LW,K,JMSW,MSW,G}.}
\end{figure}

Recent lattice results for $B$, $B_s$ and $B_c$ mesons are compiled from
\cite{LW,K,JMSW,MSW,G} in figure
\ref{bottommesons} and compared to experimental data where available.
Two groups\cite{LW,G} used the NRQCD action for the bottom quark while the
others\cite{K,JMSW,MSW} extrapolated from static quark simulations.
The light quark action
was nonperturbatively-improved clover\cite{LW,K}, twisted mass\cite{JMSW,MSW}
or HISQ (staggered)\cite{G}.

Systematic deviations of lattice from experiment are observed in some cases.
The spread among lattice results in figure \ref{bottommesons} provides an
impression of the systematic errors in current lattice simulations, but for
detailed discussions see the original lattice papers\cite{LW,K,JMSW,MSW,G}.

It should be noted that the precise and accurate lattice result for the
$B_c$ mass, coming from the HPQCD collaboration, was obtained
by tuning the standard QCD inputs to match the experimental
masses of pion, kaon, $\eta_c$ and $\Upsilon$\cite{G}.
Their $B_c^*$ mass stands as a precise prediction for future experiments.

\begin{figure}\label{staticmesons}
\includegraphics[width=11cm,trim=0 30 0 70,clip=true]{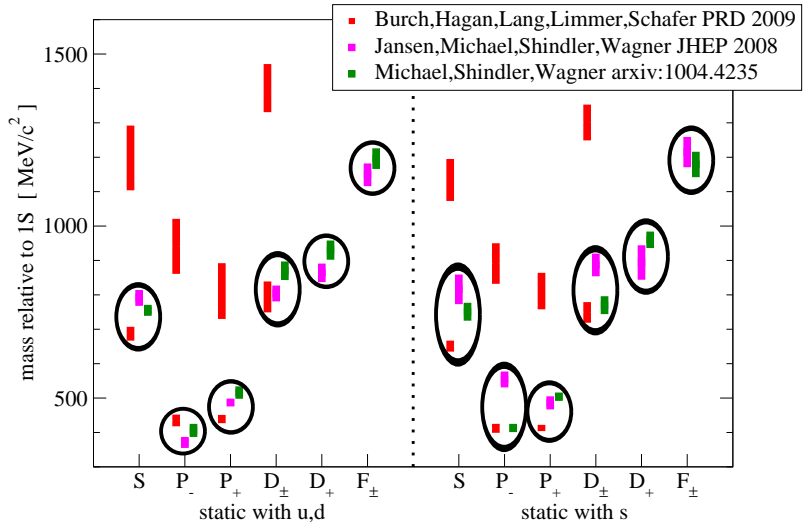}
\caption{Meson spectrum of a static valence anti-quark with a u, d or s
valence quark\cite{BHLLS,JMSW,MSW}.  Black ovals are only to guide the eye in
grouping relevant simulation results.}
\end{figure}

Interesting work has also been reported for mesons in the limit of a static
(infinitely-heavy) bottom quark\cite{BHLLS,JMSW,MSW}.
This unphysical limit shares many
qualitative features with the physical spectrum and has been used in
combination with charmed meson masses to interpolate to the region of
physical bottom mesons.  As evident from figure \ref{staticmesons}, a long
list of quantum numbers has been studied.  The variance among different
lattice determinations of some masses suggests the presence of significant
systematic effects.  Details can be found in the original
articles\cite{BHLLS,JMSW,MSW}.

\section{Charmed mesons}

\begin{figure}\label{Ds}
\includegraphics[width=7.4cm,trim=230 400 230 0,clip=true]{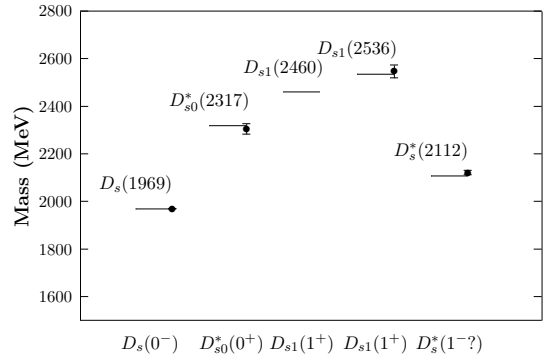}
\caption{Charmed meson spectrum from a lattice QCD simulation (filled
circles) compared to experiment (horizontal lines).  This graph was taken
directly from \cite{Kentucky}.}
\end{figure}

The spectrum of charmed mesons has not received much attention recently from
the lattice community, and thus represents a future opportunity.  Dong et
al.\cite{Kentucky}
have studied the $D_s$ spectrum, motivated in part by published speculations
that the experimentally-observed $D_{s0}^*(2317)$ might not be a $c\bar s$
state but rather a DK molecule, a four-quark state or a threshold effect.
They use domain wall sea quarks, overlap valence quarks and an Iwasaki
gauge action with a lattice spacing of 0.08 fm, a lattice volume of
(2.7 fm)$^3$ and a pion mass as light as 331 MeV.
Figure \ref{Ds} is taken directly from their conference report\cite{Kentucky}.
Dong et al.\ conclude that their preliminary results are consistent with
the $D_{s0}^*(2317)$ being a standard $c\bar s$ meson.

%%%%%%%%%%%%%%%%%%%%%%%%%%%%%%%%%%%%%%%%%%%%%%%%
%% BACKMATTER
%%%%%%%%%%%%%%%%%%%%%%%%%%%%%%%%%%%%%%%%%%%%%%%%

\begin{theacknowledgments}
Thanks to the organizers of the MENU 2010 conference, held at the College of
William and Mary, for the opportunity to participate.
This work was supported in part by the Natural Sciences and Engineering
Research Council of Canada.
\end{theacknowledgments}

\end{document}